\documentclass[12pt]{article}
\usepackage{amsmath,bbm,amssymb}
\usepackage{amsthm}
\usepackage{comment}
\usepackage{graphicx, graphics}
\usepackage{textcomp}
\usepackage{amsfonts}

\usepackage{psfrag,pst-node,subfigure,rotating, amsmath, bbm, amsthm, amssymb, amsthm, setspace, picture, epsfig, amsfonts, upgreek}

\usepackage{multirow}

\usepackage{arydshln}
\theoremstyle{definition}

\title{\bf Comparisons of penalized least squares methods by simulations}
\author{ {\Large Ke ZHANG, Fan YIN}
  \\ {\footnotesize\sl University of Science and Technology of China, Hefei 230026, China}
  \\[2mm]{\Large Shifeng XIONG\thanks{Corresponding author. Email: xiong@amss.ac.cn}}
\\ {\footnotesize\sl Academy of Mathematics and Systems Science, Chinese Academy of Sciences, Beijing 100190, China}
 }

\date{ }
\begin{document}

\maketitle \noindent{ABSTRACT:} {\normalsize \ \ \ \ Penalized least squares methods are commonly used for simultaneous estimation and variable selection in
high-dimensional linear models. In this paper we compare several prevailing methods including the lasso, nonnegative garrote, and SCAD in this area through Monte
Carlo simulations. Criterion for evaluating these methods in terms of variable selection and estimation are presented. This paper focuses on the traditional $n>p$
cases. For larger $p$, our results are still helpful to practitioners after the dimensionality is reduced by a screening method.}

\vspace{8mm}

%\vspace{8mm}
% \noindent{\normalsize {\bf 2000 Mathematics Subject Classfications}: \ \ \ \ 62J10;\ 62J15}

%\vspace{8mm} \noindent{\normalsize {\bf Running Head}: \ \ \ \ On Robust Confidence Intervals}

\newpage

\section{\bf Introduction}
\hskip\parindent

\vspace{-0.8cm}

In many fields such as business and biology, people have to deal with high-dimensional problems more and more frequently, which leads to a large demand for efficient
methods of variable selection. For high-dimensional linear regression, penalized least squares methods have been successfully developed over the last decade to
simultaneously select important variables and estimate their effects. Popular methods include the nonnegative garrote \cite{b}, the lasso \cite{t}, the elastic net
\cite{zh}, the adaptive lasso \cite{z}, SCAD \cite{fl}, and MCP \cite{z10}, among others. Theoretical properties of these methods have been actively studied. However,
there is no paper providing detailed performance comparisons between all the popular methods, which are concerned by practitioners. This paper will present such
comparisons based on numerical simulations,

\begin{comment}
Many algorithms for solving the lasso problems are available, including those developed by Fu \cite{f}, Grandvalet \cite{g}, Osborne, Presnell, and Turlach \cite{opt},
and Efron, Hastie, Johnstone, and Tibshirani \cite{ehjt}. Nowadays, the coordinate descent (CD) algorithm \cite{ts,fhht,wl,ty} is viewed as the fastest algorithm for the
lasso problem. The corresponding \texttt{R} packages include \texttt{lars} (Hastie and Efron 2011) and \texttt{glmnet} (Friedman, Hastie, and Tibshirani 2011). For the
nonconvex penalties like SCAD and MCP, existing algorithms include local quadratic approximation \cite{fl,hl}, local linear approximation \cite{zl}, the CD algorithm
\cite{bh,mfh} and the iterative soft thresholding algorithm \cite{ssw}, among others.
\end{comment}

For many applications like micro-array, one might be interested in the small $n$ and large $p$ case. For this case, Fan and Lv \cite{flv} proposed a two-stage procedure
for estimating the sparse parameter. In the first stage, a screening approach is applied to pick $M<n$ variables. In the second stage, the coefficients in the screened
$M-$submodel can be estimated by a penalized least squares method. In this paper we only focus on the traditional $n>p$ case, which can be viewed as a study on the
second stage when $p>n$. For studies on screening methods in the first stage, we refer the reader to \cite{flv,fs,hm,lpzz,w,x14}, among others.

The rest of this article is organized as follows. We describe all methods for comparison in Section \ref{sec:method}. Section \ref{sec:sim} presents the simulation
results, and Section \ref{sec:con} ends the paper with concluions.

\section{\bf Methods for comparison}\label{sec:method}
\hskip\parindent

\vspace{-0.8cm}

Consider a regression model \begin{equation}\label{lm} {Y}={X}\beta+\varepsilon, \end{equation}where ${X}=(x_{ij})$ is the $n\times p$ regression matrix,
$Y=(Y_1,\ldots,Y_n)\in{\mathbb{R}}^n$ is the response vector, $\beta=(\beta_1,\ldots,\beta_p)'$ is the vector of regression coefficients, and $\varepsilon$ is the vector
of random errors with mean 0 and variance $\sigma^2<\infty$. We assume that there is no intercept, which holds when $X$ is standardized as $\sum_{i=1}^{n}x_{ij}=0$ and
$\sum_{i=1}^{n}x_{ij}^2=n$ and $Y$ is centered as $\sum_{i=1}^nY_i=0$.

\subsection{Ordinary least square}

The ordinary least square (OLS) method is a basic approach to estimate $\beta$. Its expression is given by
\begin{equation*}
\hat{\beta}_{\mathrm{OLS}}= (X'X)^{-1}X'Y
\end{equation*}
The OLS estimator is widely used and can serve as the initial estimator in many other methods such as the nonnegative garrote and adaptive lasso. In our simulations, we
use the function "lm" in \texttt{R} to compute the OLS estimator.

\subsection{Ridge regression}

Ridge regression \cite{hk} uses an $\ell_2$-norm penalty to improve OLS when the covariates are correlated. Like OLS, the ridge estimator has an explicit form
\begin{equation}
\hat{\beta}_{\mathrm{ridge}} = (X'X+\lambda I_{p})^{-1}X'Y,\label{ridge}
\end{equation}
where $\lambda\geqslant0$ is the tuning parameter and $I_p$ denotes the $p\times p$ identical matrix. Here we select $\lambda$ by minimizing the generalized
cross-validation criterion (GCV) \cite{ghw}
\begin{equation}
\mathrm{GCV}(\lambda) = \frac{||(I_n-A(\lambda))y||^2}{({\mathrm{Trace}}(I_n-A(\lambda)))^2},\label{gcv}
\end{equation}where $A(\lambda)=X(X'X+\lambda I_{p})^{-1}X'$.
%Using the strategy similar to \cite{7}, 100 value of tuning parameters $\lambda$ should be place on the log scale. Then we could choose the $\lambda_{min}$ which
%minimize the V($\lambda$) to compute the $\hat{\beta}_{ridge}$ in equation(4).

\subsection{The nonnegative garrote}

The nonnegative garrote (NG) method \cite{b} is a direct shrinkage of OLS through multiplying it by a nonnegative factor $u=(u_1,...,u_p)'$. The NG estimator has the
form $\hat{\beta}_{\mathrm{NG}}=(u_1\hat{\beta}_1,\ldots,u_p\hat{\beta}_p)'$, where $u$ is the solution to the convex quadratic optimization problem
\begin{eqnarray}
\min_{u\geqslant0}\big\{||Y-X\hat{B}u||^2+2\lambda\sum_{j=1}^{p} u_j\big\} \label{ng}
\end{eqnarray} and $\hat{B}={\mathrm{diag}}(\hat{\beta_1},...,\hat{\beta_p})$, $(\hat{\beta}_1,\ldots,\hat{\beta}_p)'$ is the OLS estimator,
and $\lambda>0$ is the tuning parameter.

Xiong \cite{x} showed that the NG estimator can be obtained by directly minimizing some model selection criteria such as Mallows's $C_p$, AIC, and BIC. Based on this,
$\lambda$ in \eqref{ng} can be accordingly chosen as $\hat{\sigma}^2$ ($C_p$ and AIC) or $\hat{\sigma}^2\log{n}/2$ (BIC), where $\hat{\sigma}^2$ is the standard
estimator of $\sigma^2$ based on OLS.

When the covariates are highly correlated, we can use the ridge estimator in \eqref{ridge} as the initial estimator in NG to improve the initial NG \cite{yl,x}. Xiong
\cite{x} proposed the following ridge-based NG estimator $\hat{\beta}_{\mathrm{rNG}}=(u_1\tilde{\beta}_1,\ldots,u_p\tilde{\beta}_p)'$, where
$\tilde{\beta}=(\tilde{\beta}_1,\ldots,\tilde{\beta}_p)'$ is the ridge estimator with tuning parameter $\lambda_{\mathrm{r}}$ derived from minimizing GCV \eqref{gcv},
$u$ is the solution to
\begin{equation}\min_{u\geqslant0}\big\{\|Y-{X}\tilde{B}u\|^2+2\lambda\sum_{j=1}^pw_ju_j\big\},\label{nr}\end{equation}and
$\tilde{B}={\mathrm{diag}}(\tilde{\beta_1},...,\tilde{\beta_p})$. Here $w_j$ in \eqref{nr} is the $(j,j)$ entry of matrix
$({X}'{X}+\lambda_{\mathrm{r}}{I}_p)^{-1}({X}'{X})$. The selection of $\lambda$ in \eqref{nr} is the same as the initial NG method above according to \cite{x}, i.e.,
$\hat{\sigma}^2$ or $\hat{\sigma}^2\log{n}/2$. In our simulations, the function ``quadprog" in \texttt{matlab} is used to compute the shrinkage factor $u$ in \eqref{ng}
and \eqref{nr}.

\subsection{The lasso and elastic net}
The lasso \cite{t} is a method which assigns an $\ell_{1}$ penalty to the model and get a sparse solution. The elastic net \cite{zh} uses the mixture of $\ell_{1}$ and
$\ell_{2}$ penalty to improve it when the covariates are correlated. Specifically, the elastic net estimatro is the solution to
\begin{equation}
\min_{\beta}\big\{\|Y-X\beta\|^2+2\lambda_1\sum_{j=1}^p|\beta_j|+\lambda_2\sum_{j=1}^p\beta_j^2\big\},
\end{equation}
where $\lambda_1$ and $\lambda_2$ are nonnegative tuning parameters. When $\lambda_2=0$, the above estimator reduces to the lasso. The two tuning parameters can be
selected by cross-validation. In our simulations, to reduce the computational intensity, we set $\lambda_1 = \lambda_2$ for elastic net and we use \texttt{R} package
\texttt{glmnet}, which is based on the coordinate descent algorithm \cite{fhht}, to implement the lasso and elastic net.

\subsection{The adaptive lasso}

The adaptive lasso \cite{z} solves the problem
\begin{equation}
\min_{\beta}\big\{||Y-X\beta||^2+2\lambda\sum_{j=1}^{p} \hat{w}_j\beta_j\big\},
\end{equation}
where the weights $\hat{w}_j$'s are added for reducing the bias of the lasso.

In our comparison, we use \texttt{R} package \texttt{parcor} \cite{nj} to compute the adaptive lasso. In this package, they set the weight factor as a function of the
lasso estimator: $$\hat{w_i}=\frac{1}{|\hat{\beta}_{i}|},$$where $\hat{\beta}=(\hat{\beta}_1,\ldots,\hat{\beta}_p)'$ is the lasso estimator with the tuning parameter
derived from 10-fold cross-validation.

\par Like lasso, the tuning parameter $\lambda$ of adalasso is chosen by 10-fold cross-validation. However, as \cite{nj} puts, In each of the k-fold
cross-validation steps, the weights for adaptive lasso are computed in terms of a lasso fit. This implies that a lasso solution is computational expensive.

\subsection{SCAD and MCP}

SCAD \cite{fl} and MCP \cite{z10} are two penalized methods with nonconvex penalties. They are the solutions to
\begin{equation}
\min_{\beta}\big\{||Y-X\beta||^2+P_{\lambda,\gamma}(\beta)\big\},
\end{equation}where\begin{equation*}
P_{\lambda,\gamma}'(t) =  \lambda\left\{I(t<\lambda) + \frac{(\gamma\lambda-t)_{+}}{(\gamma-1)\lambda} I(t>\lambda) \right\}
\end{equation*}for SCAD and\begin{equation*}
P_{\lambda,\gamma}'(t) = \left(\lambda-\frac{t}{\gamma}\right)_{+}
\end{equation*} for MCP.

In our simulation, we use \texttt{R} package \texttt{ncvreg} to implement SCAD and MCP. This package was developed by \cite{bh} based on the coordinate descent algorithm
\cite{bh,mfh}. The tuning parameters $\lambda$ and $\gamma$ are chosen as follows \cite{bh}: BIC and
convexity diagnostics are used to choose an appropriate value of $\gamma$ and ten-fold cross-validation was then used to choose $\lambda$ for MCP and SCAD.

\section{\bf Simulations}\label{sec:sim}
\hskip\parindent

\vspace{-0.8cm}

In our simulations, we generate data from the model $$Y_i=\beta_0+\beta'x_i+\varepsilon_i$$ for $i=1,\ldots,n$ with 1000 repetition times, where
$\varepsilon_1,\ldots,\varepsilon_n$ are i.i.d. from $N(0,\sigma^2)$. The vectors $x_1,...,x_n$ are i.i.d. from $N(0,\Sigma)$, where the $(i,j)$ entry of $\Sigma$ is
$\rho^{|i-j|}$.

In this section, all the pictures and charts are exhibited from the data generated from above model. We use the methods in Section \ref{sec:method} to estimate $\beta$.
For an estimator $\hat{\beta}=(\hat{\beta}_1,\ldots.\hat{\beta}_p)'$, we compute the mean squared error (MSE) $E\|\hat{\beta}-\beta\|^2$, the model error (ME)
$E(\hat{\beta}-\beta)'X'X(\hat{\beta}-\beta)$ to show the estimation and prediction performance. We also compute the first kind of incorrect number (IC1) and the second
kind of incorrect number (IC2) to show the accuracy of variable selection, where
\begin{eqnarray*}
IC1 = \#\{j:\beta_j\neq 0 , \hat{\beta}_{j}=0\} \notag \\
IC2 = \#\{j:\beta_j= 0 , \hat{\beta}_{j}\neq 0\} \notag.
\end{eqnarray*} All the simulations are implemented via intel core i3-380 (2.53GHz).

\subsection{Different correlations}

\begin{figure}[htp]
  \includegraphics[width=0.55\textwidth,angle=0]{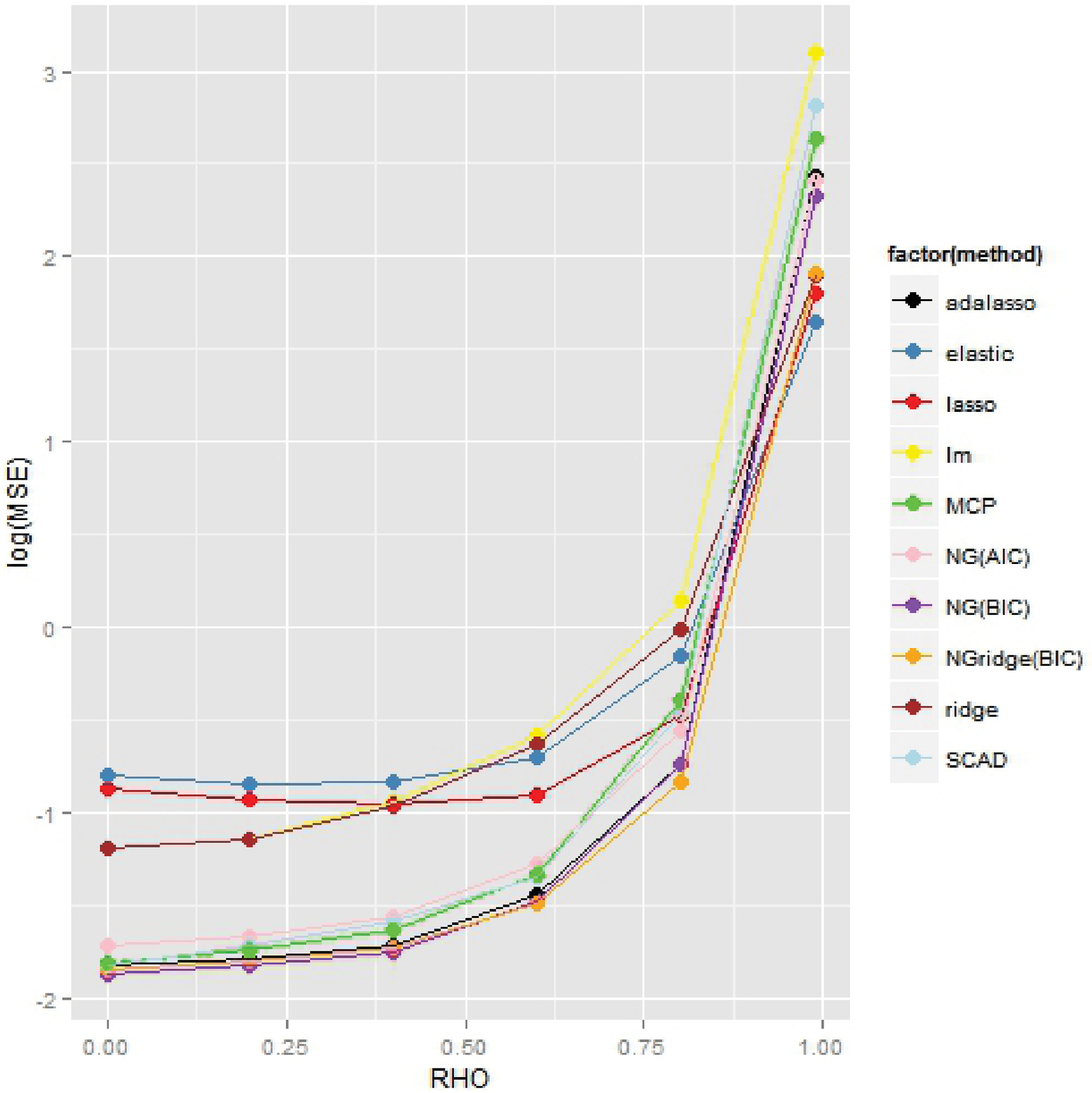}
  \includegraphics[width=0.55\textwidth,angle=0]{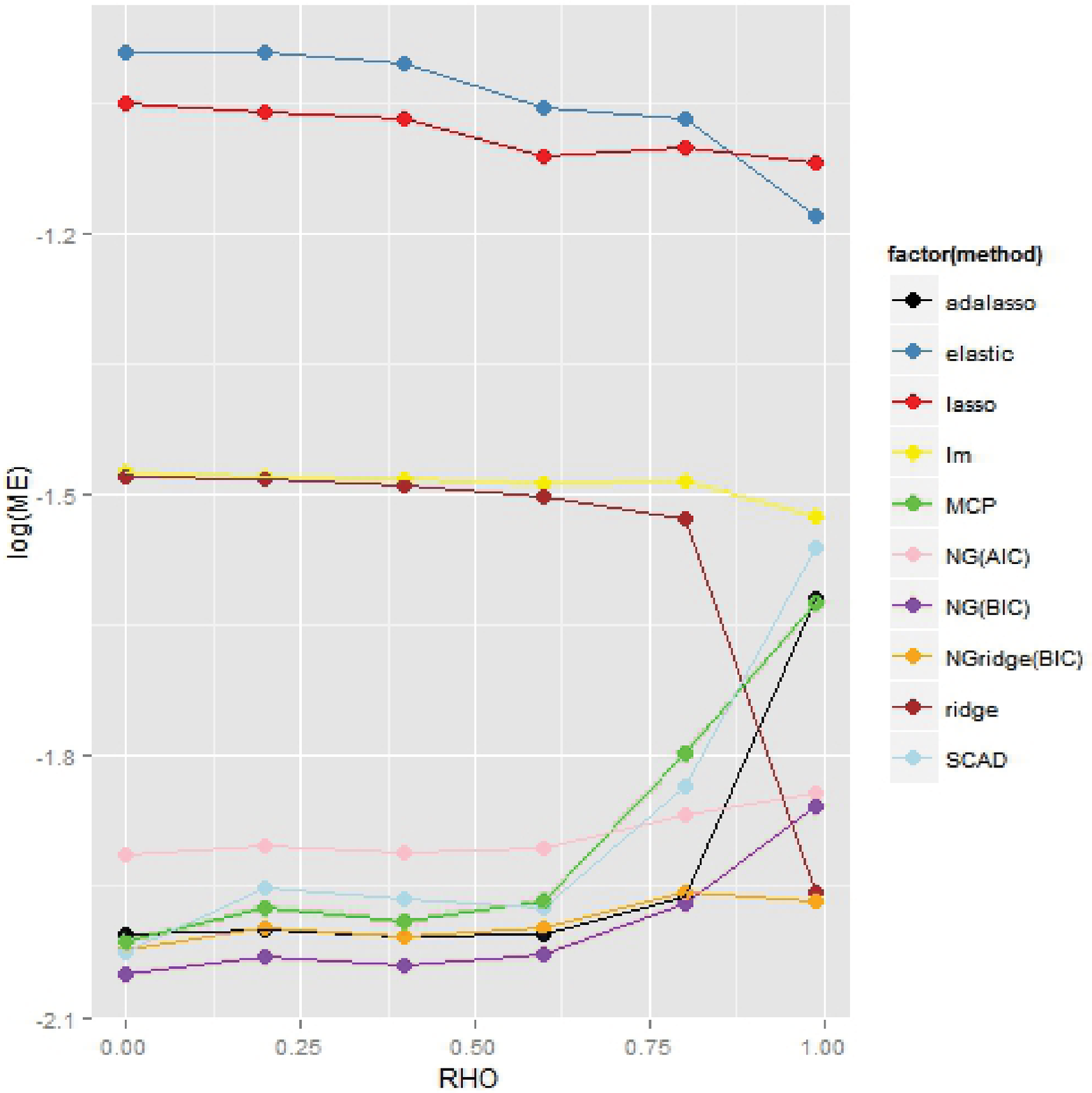}
  \includegraphics[width=0.55\textwidth,angle=0]{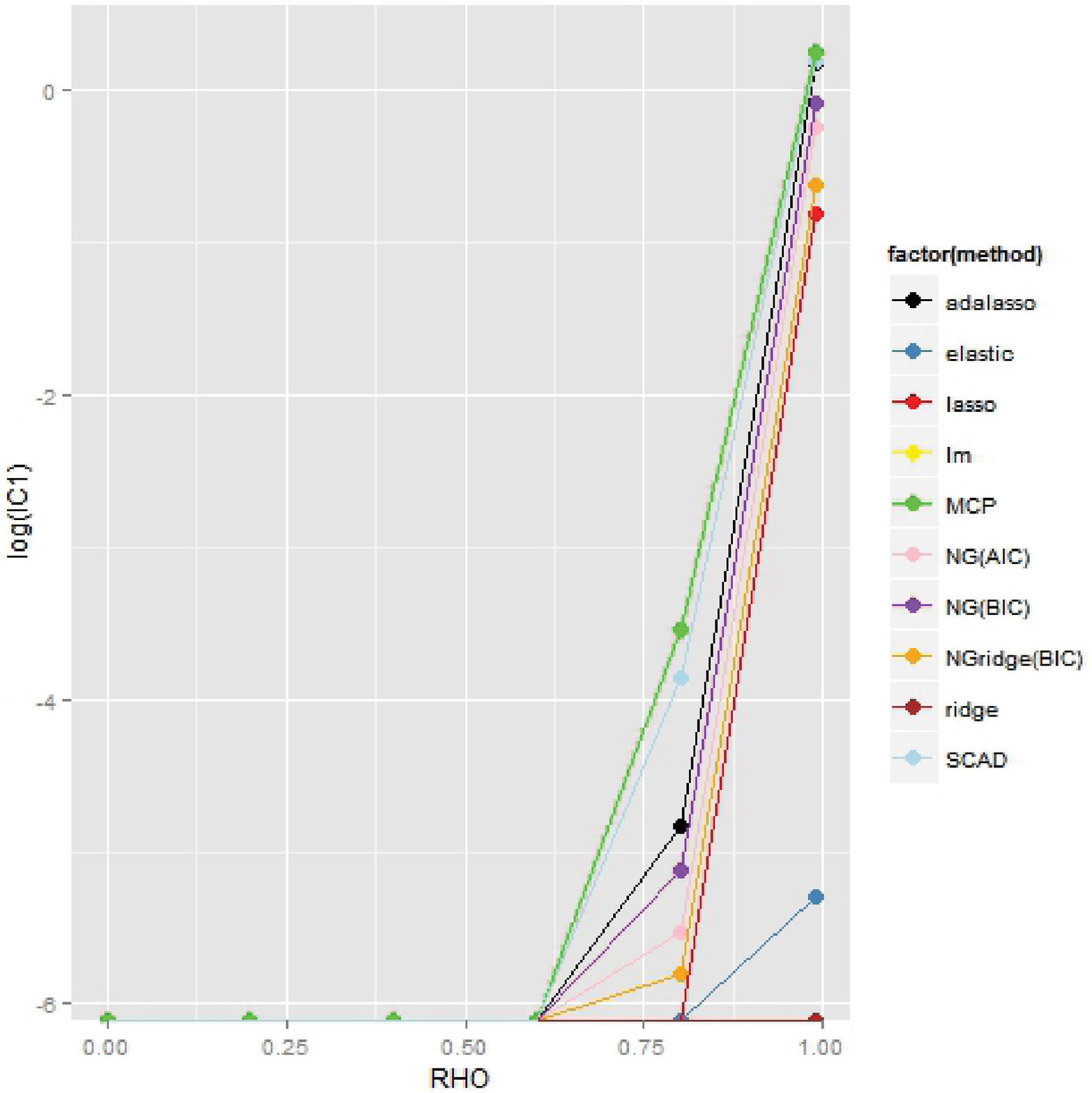}
  \includegraphics[width=0.55\textwidth,angle=0]{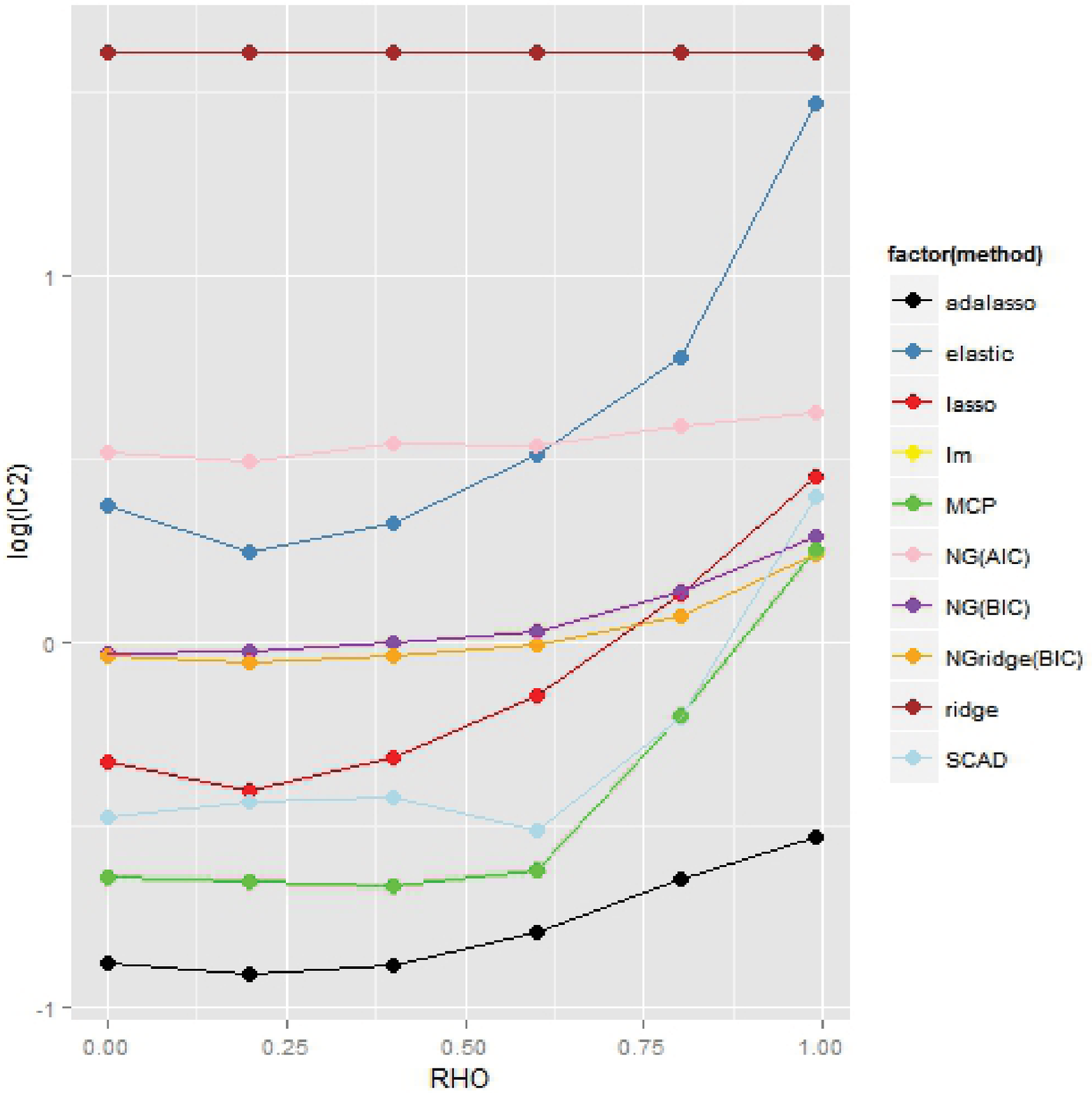}\caption{$n=40,p=8,\sigma=1$}
\label{fig:sig1}
 \end{figure}

\begin{figure}[htp]
  \includegraphics[width=0.55\textwidth,angle=0]{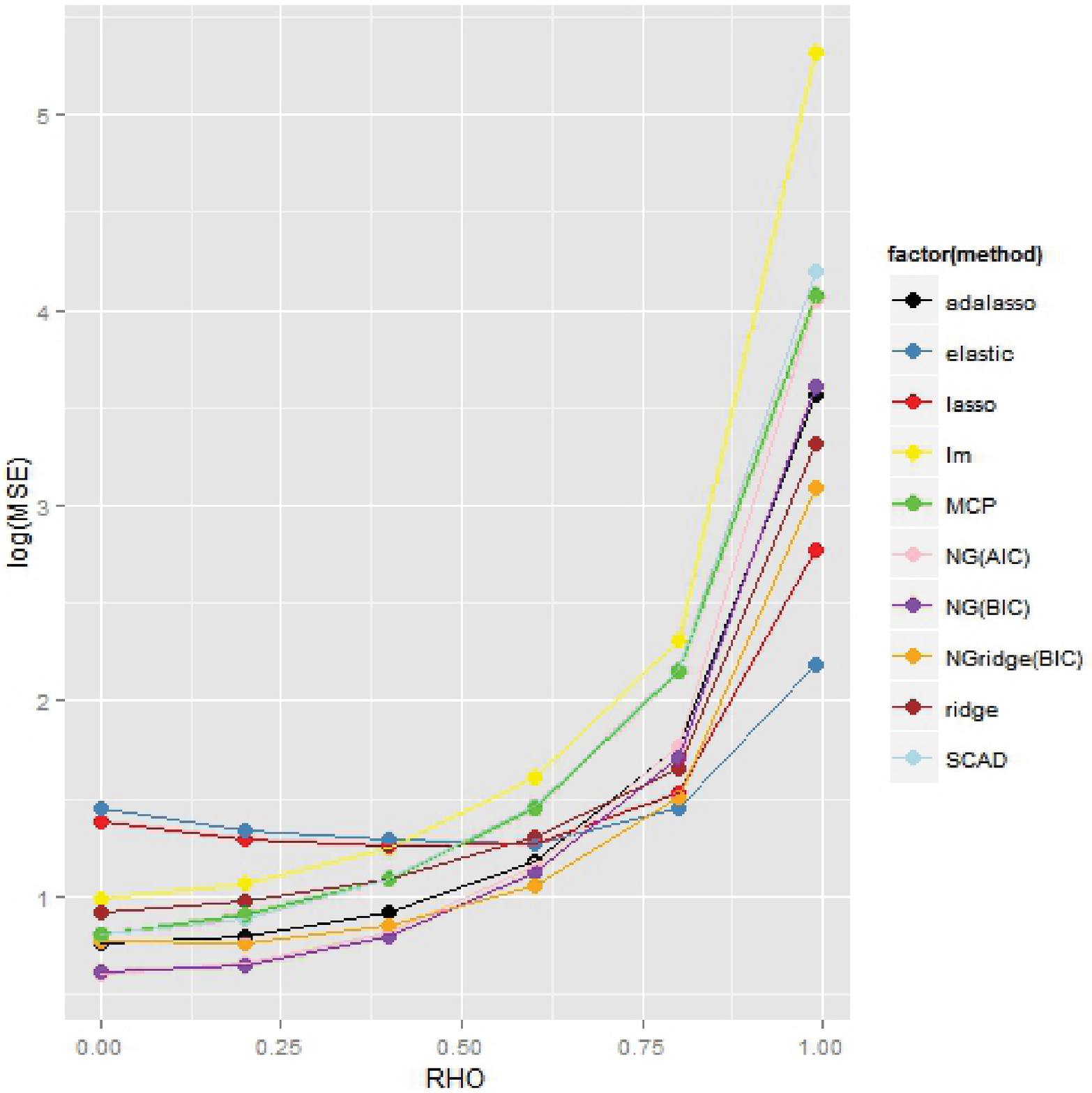}
  \includegraphics[width=0.55\textwidth,angle=0]{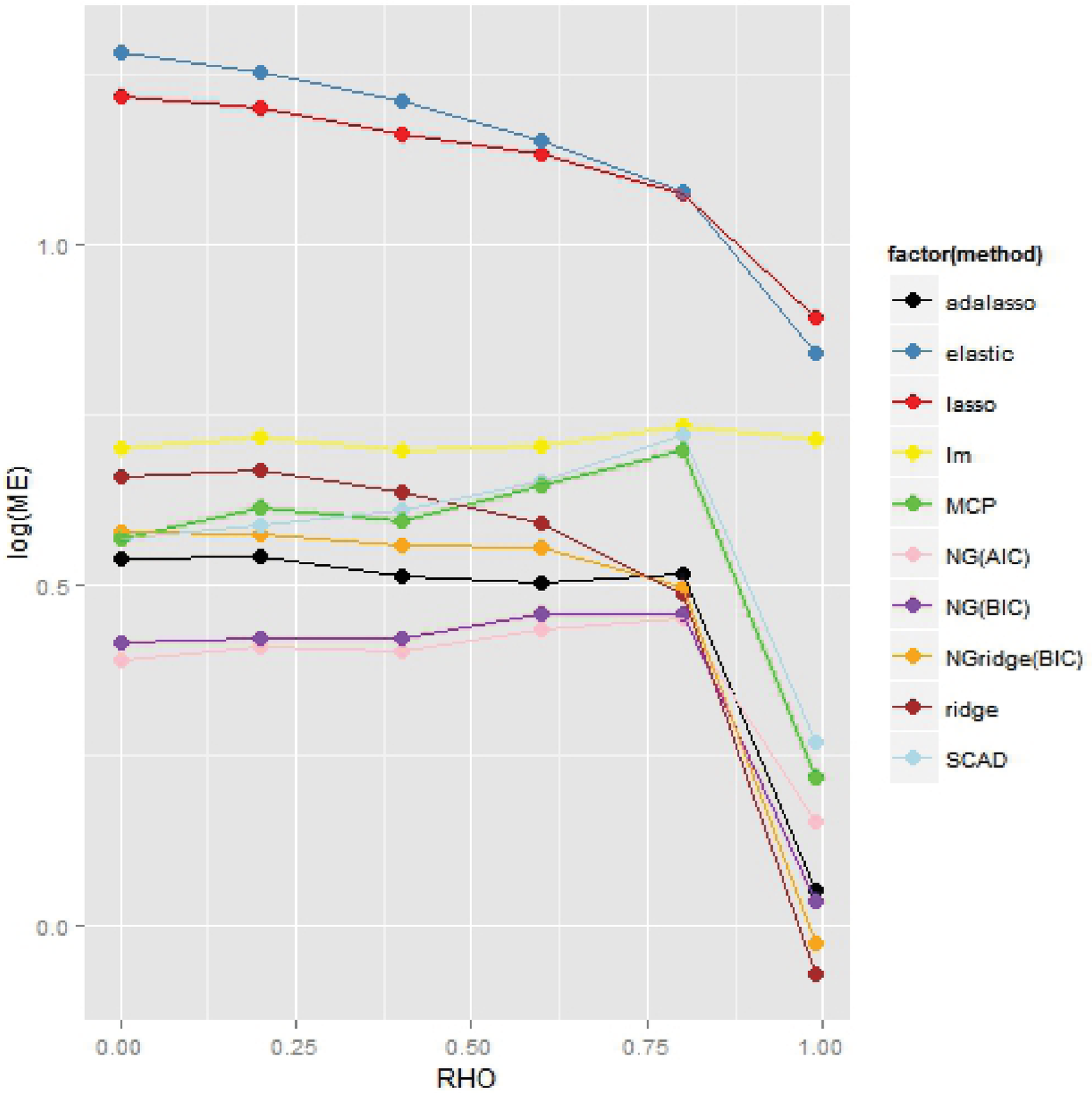}
  \includegraphics[width=0.55\textwidth,angle=0]{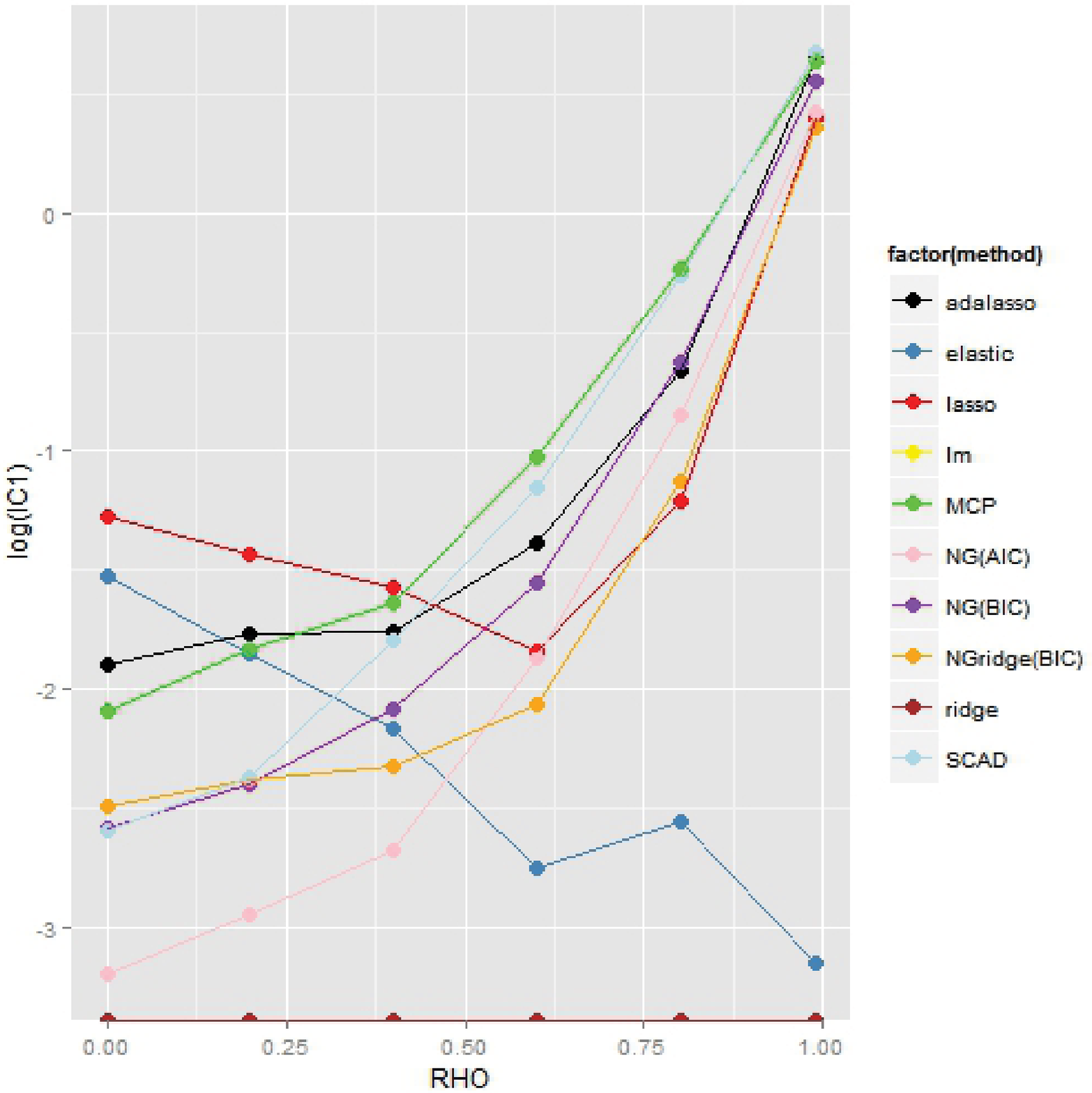}
  \includegraphics[width=0.55\textwidth,angle=0]{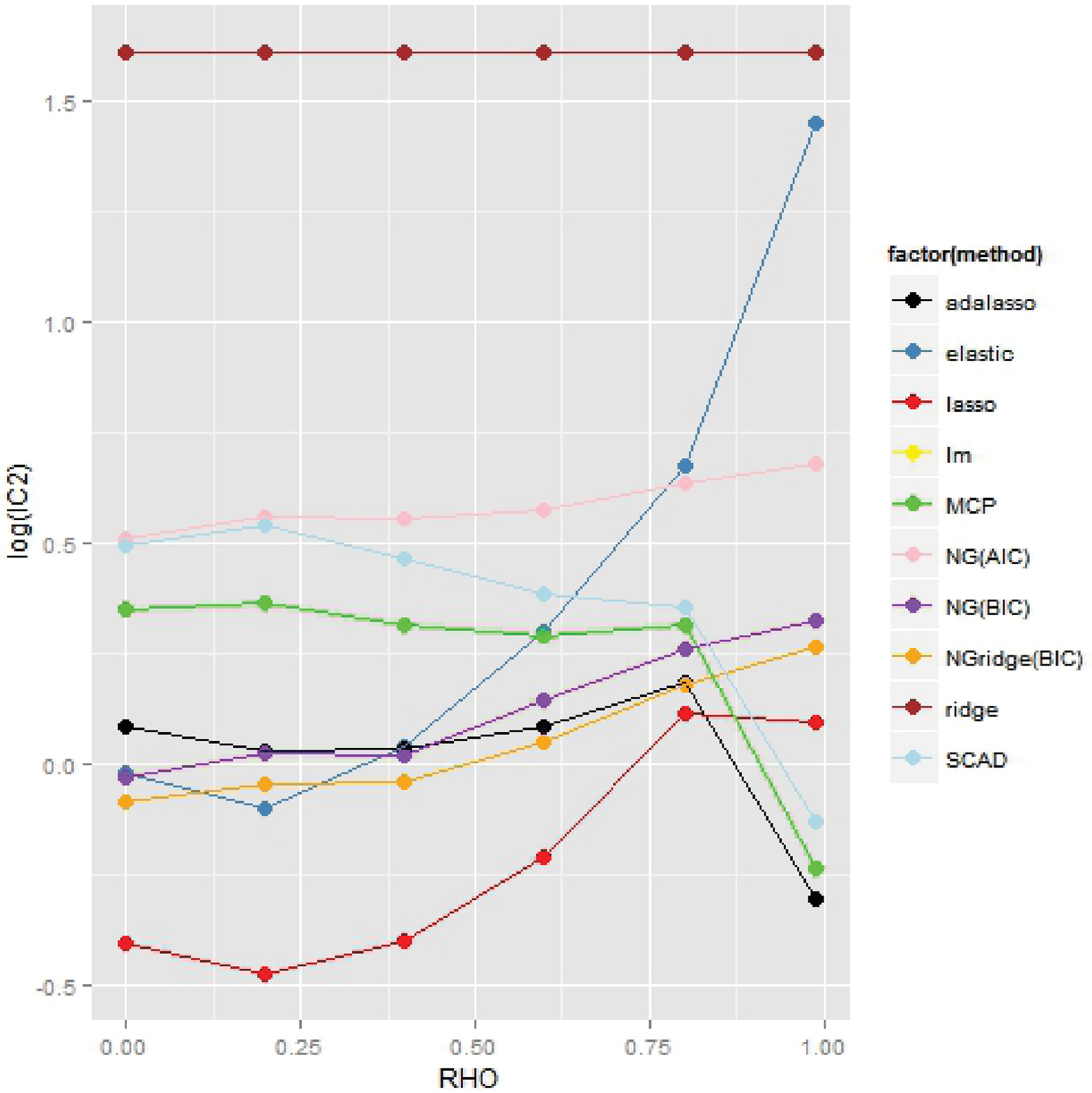}\caption{$n=40,p=8,\sigma=3$}
\label{fig:sig3}
 \end{figure}

\begin{figure}[htp]
  \includegraphics[width=0.55\textwidth,angle=0]{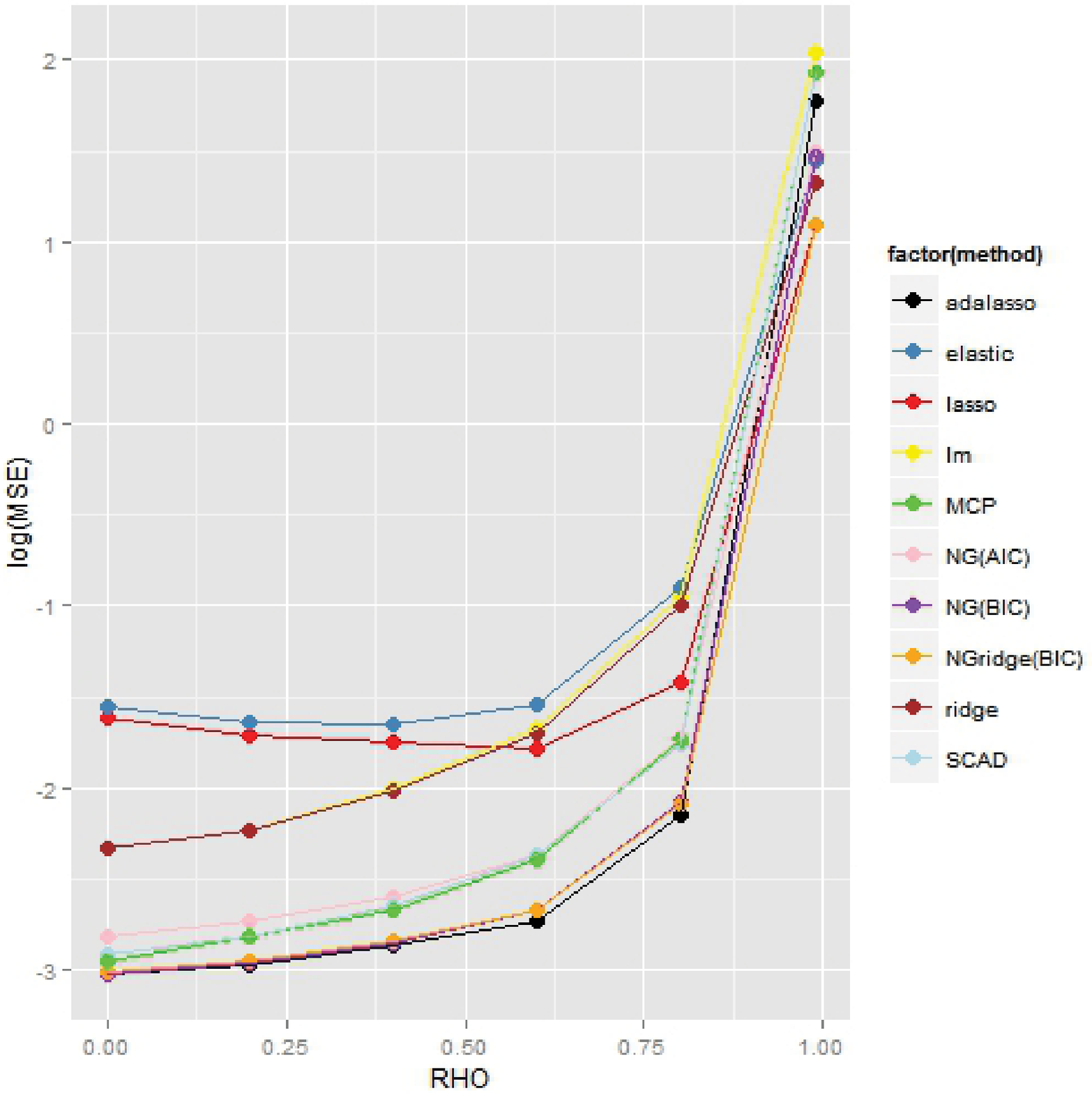}
  \includegraphics[width=0.55\textwidth,angle=0]{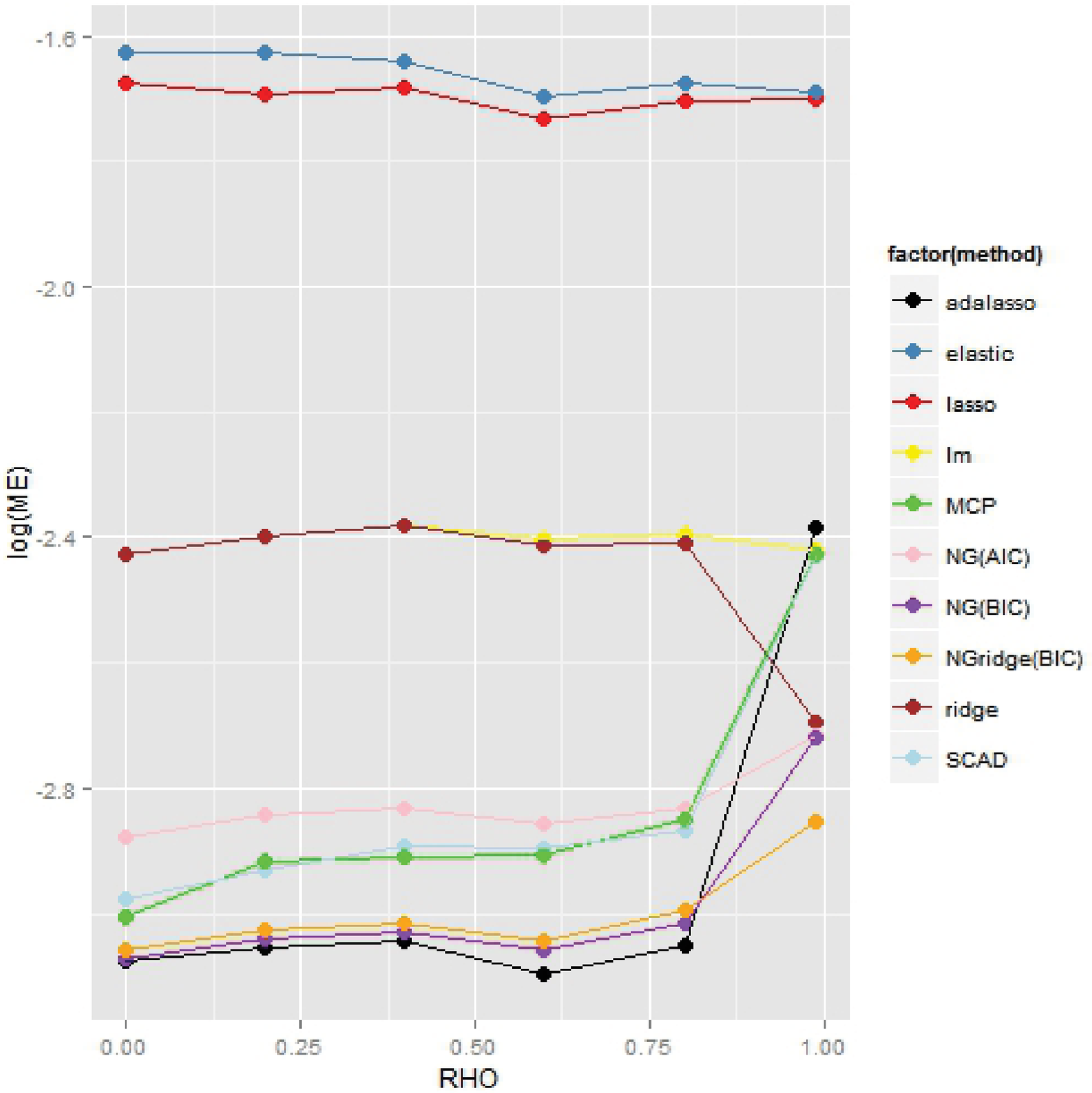}
  \includegraphics[width=0.55\textwidth,angle=0]{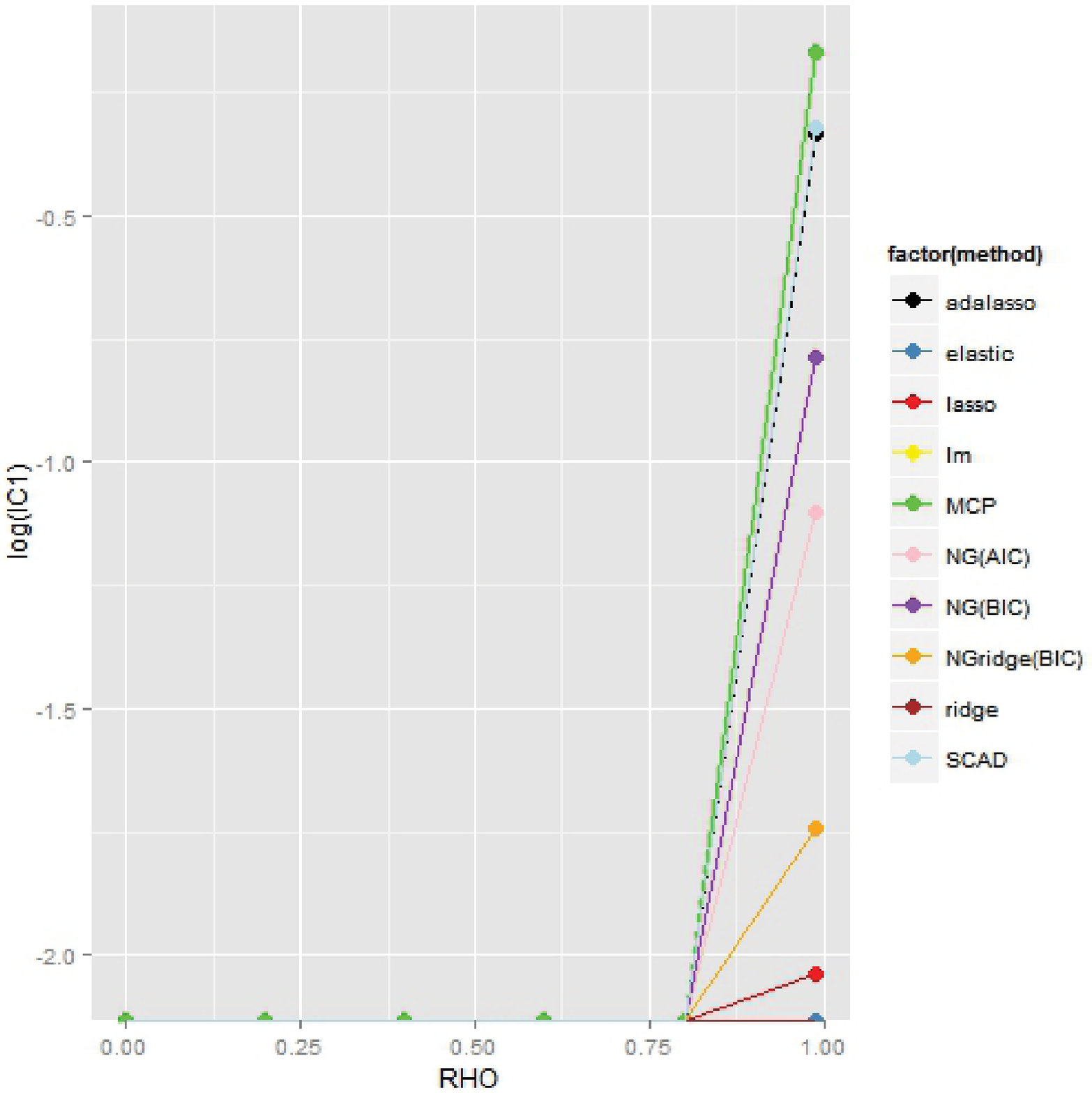}
  \includegraphics[width=0.55\textwidth,angle=0]{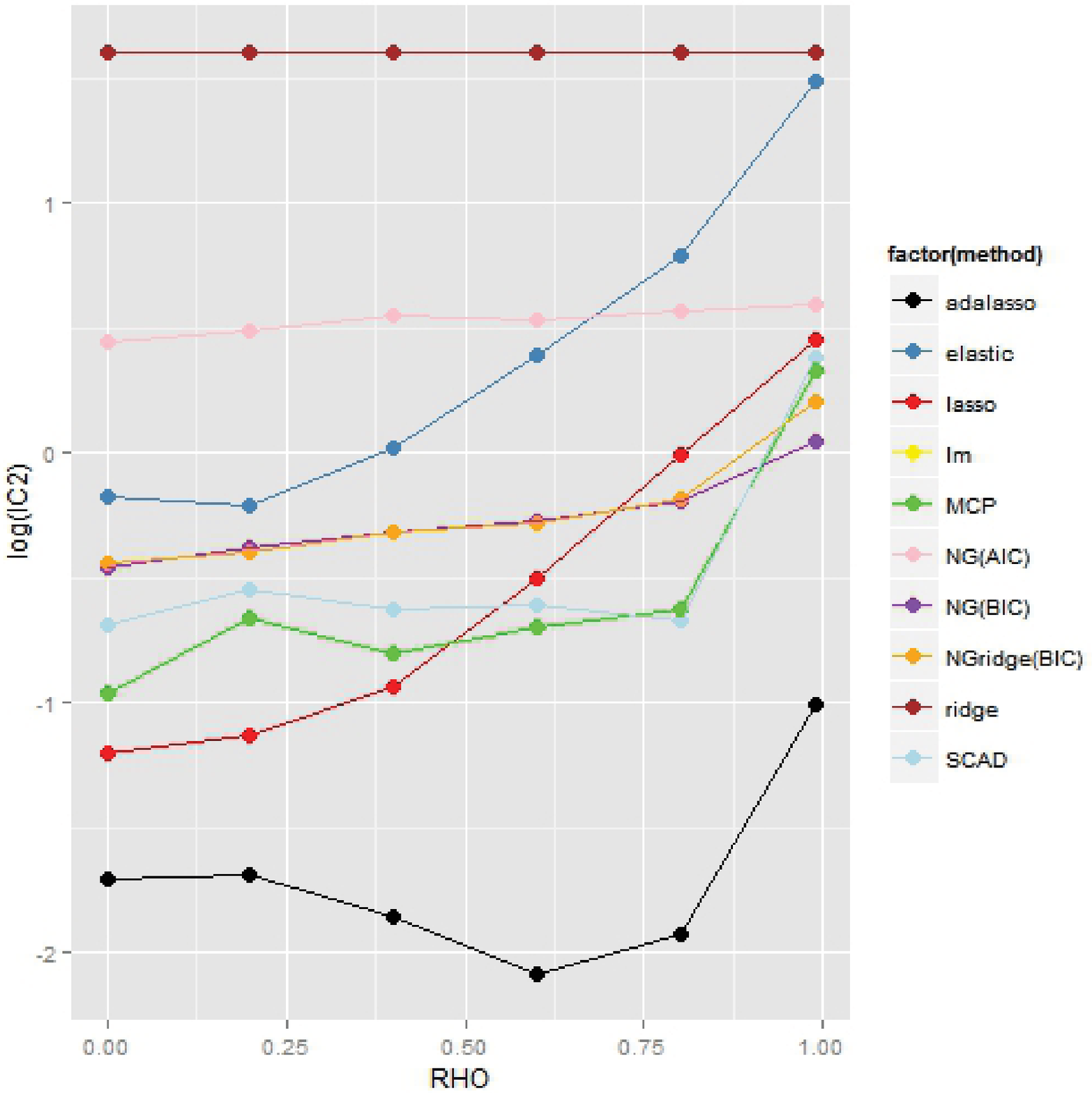}
  \caption{$n=100,p=8,\sigma=1$}
\label{fig:n100}
 \end{figure}

In this subsection, we fix $\beta_0=4$, and let the true coefficient vector $\beta$ be $(3,1.5,0,0,2,0,0,0)$. The situations where the correlation parameter $\rho$
varies from 0 to 0.99 are considered for three configurations of $(n,\sigma)$: Case I: $n=40,\sigma=1$; Case II: $n=40,\sigma=3$; Case III: $n=100,\sigma=1$. The
corresponding results are shown in Figures \ref{fig:sig1}-\ref{fig:n100}. To make these figures easier to observe, we apply the log transformation to the $y$ axes.

To begin with, the pictures show that for most $\rho$, the methods with ability of variable selective perform better than OLS and ridge. However the lasso and elastic
net not work well when the $\rho$ is relatively lower. They always have comparably higher ME, meaning that the prediction of training data is not so accurate.

Then we will alter the condition by making the noise greater or setting more training data. We could observe that the adalasso shows the best accurate of variable
selection(lower IC2) when we have enough training data. On the other hand, as a compromise of lasso and ridge, elastic net behaves conservatively and tent to reduce IC1
and thus increase IC2.

\subsection{Nearly sparse models}

\begin{figure}[htp]
  \includegraphics[width=0.55\textwidth,angle=0]{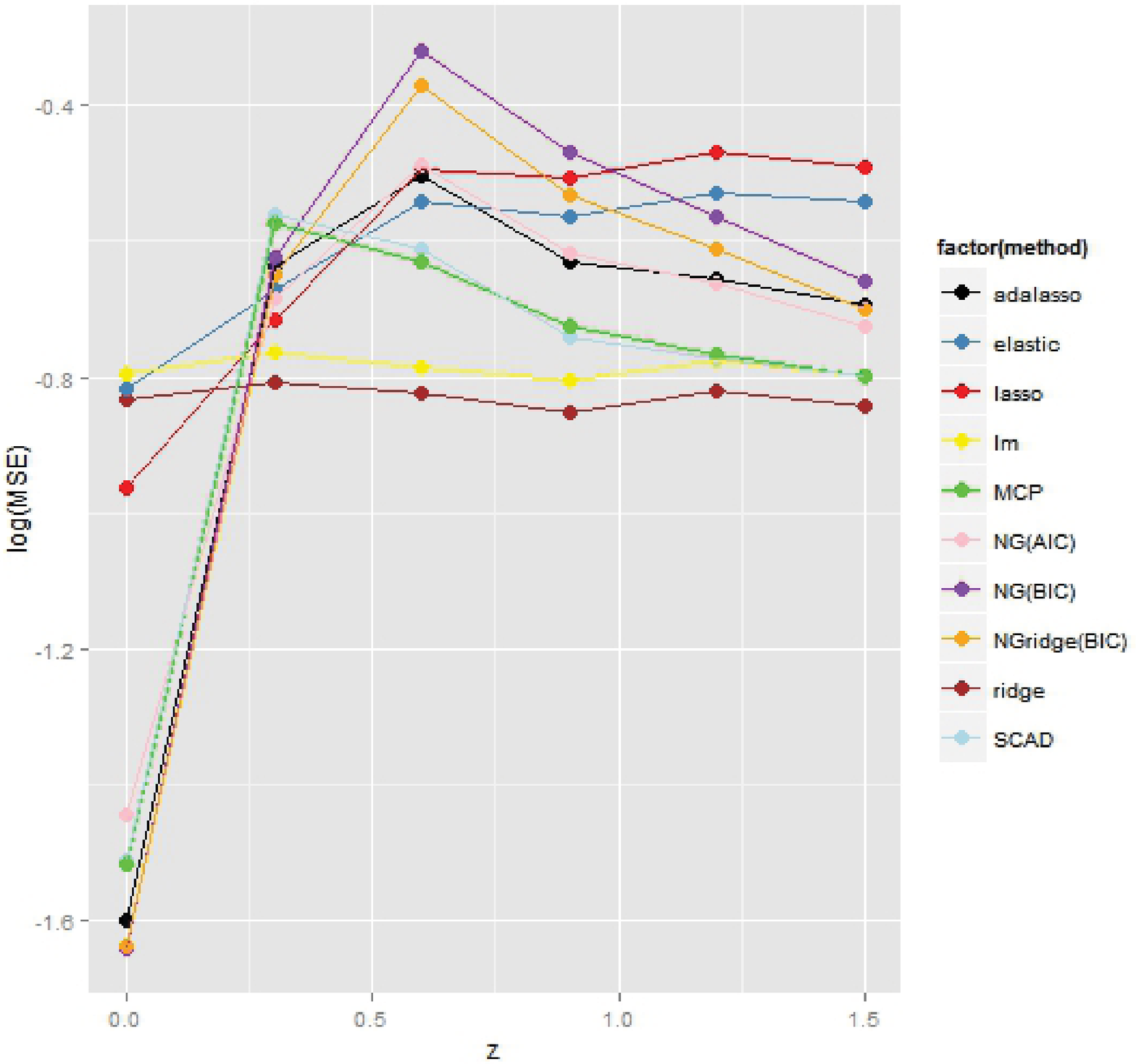}
  \includegraphics[width=0.55\textwidth,angle=0]{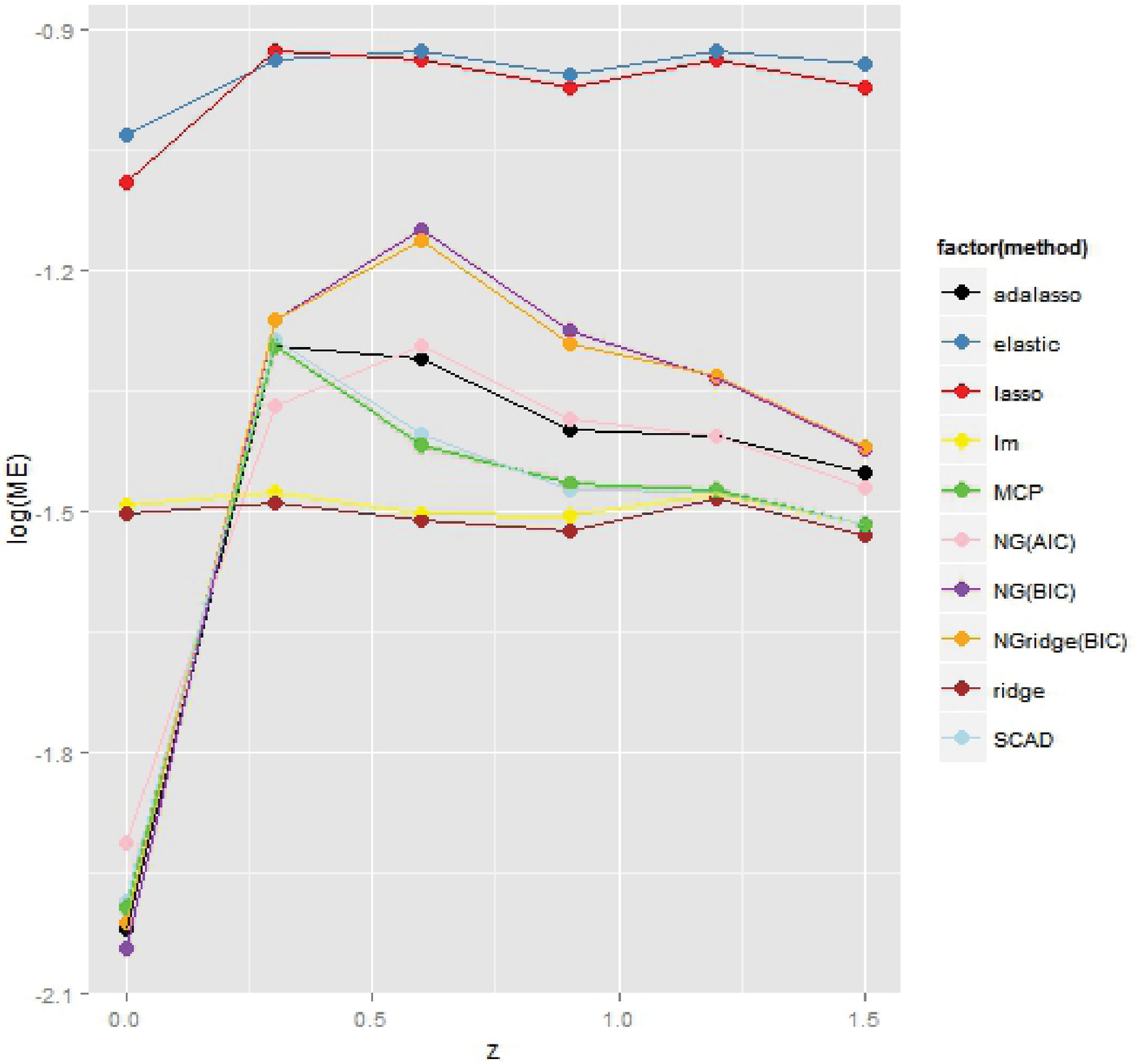}\caption{$n=40,p=8,\sigma=1,\rho=0.5$}
\label{fig:vb}
\end{figure}

We next consider the performance of these methods when the coefficients are not sparse but some of them are close to 0. In this subsection, $n=1$, $\sigma=1$,
$\rho=0.5$, and $\beta$ is set to $(4,3,1.5,z,z,2,z,z,z)$, where $z$ varies from 0 to 1.5. The results are shown in Figure \ref{fig:vb}.

According to the result above, in this simulation, there is a turning point for some variable selective methods. Firstly, for OLS and ridge, it is seemingly that they
are not sensitive to the changing of parameter and they outperform when $z$ is larger than 0.3. On the other hand, for SCAD, MCP and adaptive lasso, the turning point is
approximately 0.3 while for the three kinds of NG, the turning point is about 0.6. However, without obvious turning point, the lasso and elastic net don't perform well
in this approximate sparse model.

\subsection{Higher dimensions}

\begin{figure}[htp]
  \includegraphics[width=0.55\textwidth,angle=0]{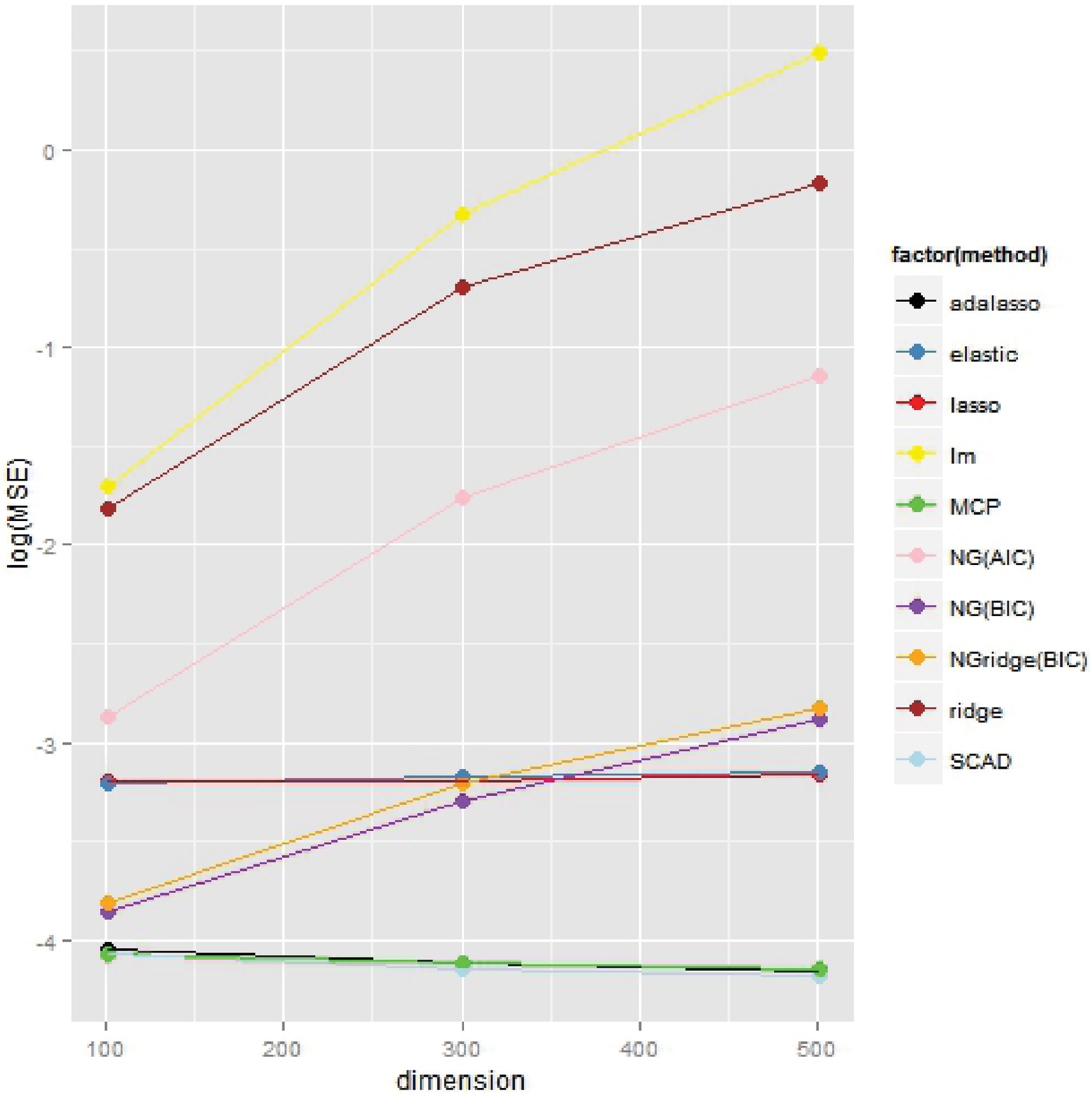}
  \includegraphics[width=0.55\textwidth,angle=0]{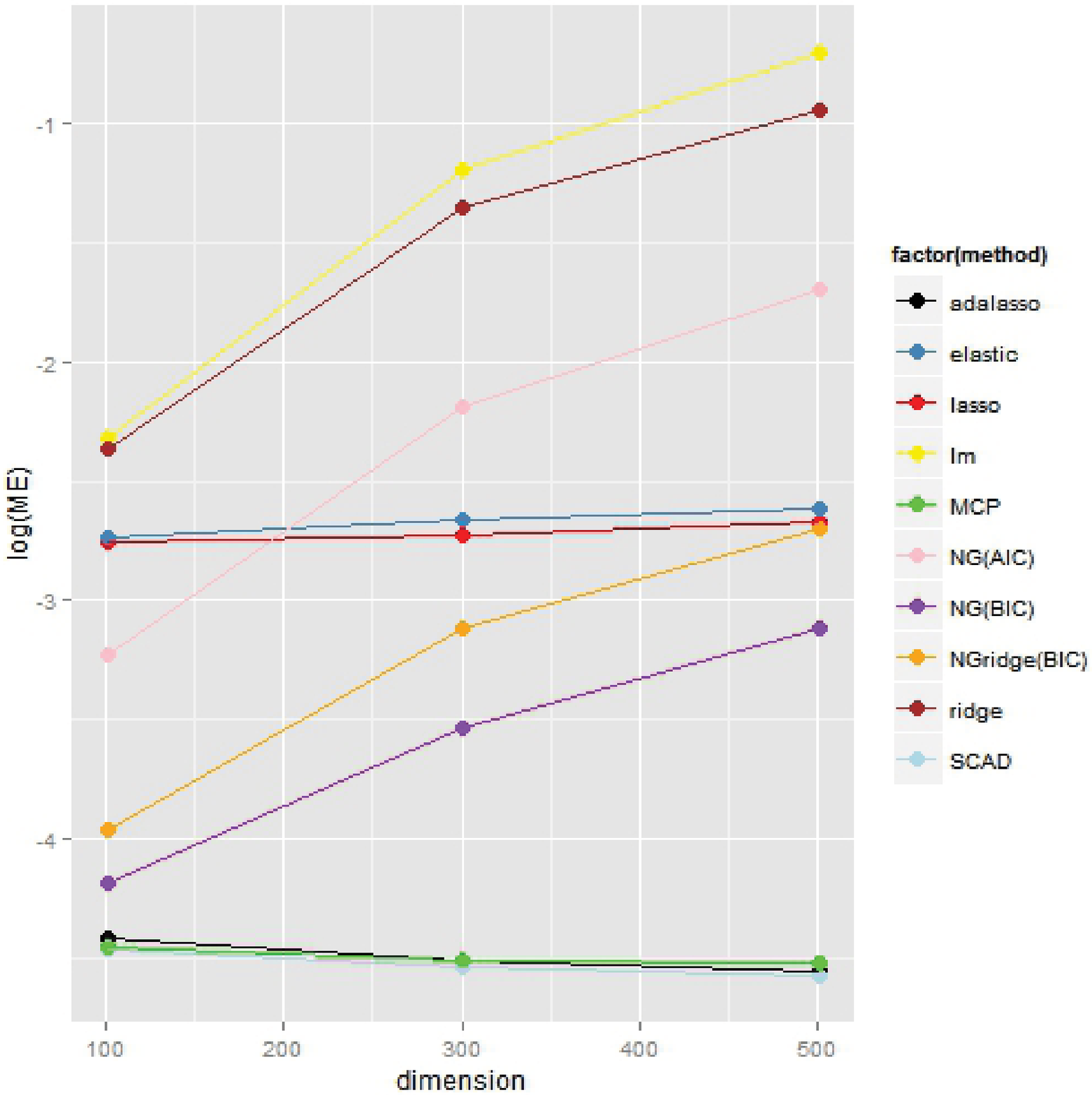}\caption{$n=1000,\sigma=1,\rho=0.5$}
\label{fig:vp}
\end{figure}

This subsection compares the performance of these methods for larger $p$. Here $n$ is set to be 1000 and we fixed the $\rho$ and $\sigma$ to 0.5 and 1, respectively.
Then we simulate with $p$ varying from 100 to 300 for $\beta= (1,\ldots,1,0,\ldots,0)$, where the nonzero components in $\beta$ is 10. The simulation results are
displayed in Figure \ref{fig:vp}.

\par According to the pictures above, adalasso, SCAD and MCP have advantages overweighing those of other methods in this condition, and they show a good tendency
to be more accurate if $p$ increase. However the three kind of NG seems to obtain higher MSE and ME.

\par Because of the relatively larger amount of data size, almost none of the methods make the first kind of mistake (IC1), which is not presented in the picture above.
Moreover, the three kinds of NG may have difficulty finding the real parameter (higher IC2) when the scale of dimension gets higher.

\subsection{Time Cost comparison}

\begin{figure}[htp]
  \includegraphics[width=1\textwidth,angle=0]{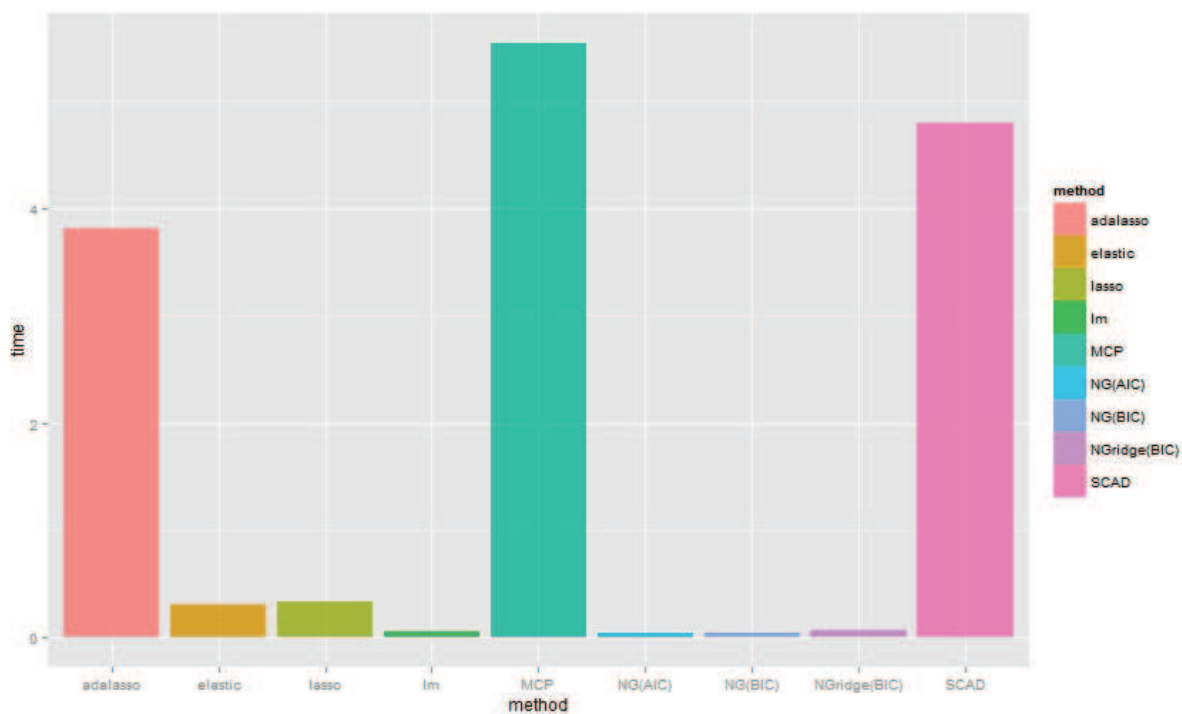}\caption{$n=1000,p=100,\sigma=1,\rho=0.5$}
\label{fig:t}
\end{figure}

To compare the computational time of these methods, we simulate for $n=1000$, $\rho=0.5$, and $p=100$. The results for time comparison are shown in Figure \ref{fig:t}.

\section{\bf Conclusions}\label{sec:con}
\hskip\parindent

\vspace{-0.8cm}

In summary, according to the result of our simulation, we attempt to itemize the properties of all the methods:

\begin{itemize}
\item \textbf{OLS and Ridge}
\newline OLS and Ridge are the basic methods without the ability of choosing parameters, they are the methods which do not perform well in our simulation except
the approximate sparse model.

\item \textbf{Lasso and Elastic Net}
\newline Lasso and Elastic Net always find it diffecult predicting parameters accurate when $\rho$ is relatively lower. Elastic net present a conservative estimate
with lower IC1 and higher IC2.

\item \textbf{MCP and SCAD}
\newline MCP and SCAD perform well in high dimension condition and tend to estimate with higher IC1 and lower IC2.

\item \textbf{Adaptive Lasso}
\newline With enough training data(oracle property) and relatively lower noise, adalasso could often select variable accurately. However, it is computational intensive
and do not work well with extreme correlation.

\item \textbf{Nonegative Garrote(NG)}
\newline Three kinds of NG works well with relatively greater noise, but their variable selective ability are damaged when the dimension gets higher.
On the other hand, it is obvious that NG(BIC) and NGridge(BIC) are better than NG(AIC).
\end{itemize}

\vspace{1cm} \noindent{\bf Acknowledgements}

Xiong's research is supported partly by the National Natural Science Foundation of China (Grant No. 11271355).

%\vspace{1cm}
\bibliographystyle{model1a-num-names}
\bibliography{<your-bib-database>}
 
\end{document}